\documentclass[11pt]{article}

\usepackage{amssymb}
\usepackage{epsfig}

\def\dd{\displaystyle}

\def\aa{\alpha}

\def\dd{\displaystyle}

\def\xj{x_t^j}
\def\xit{x_t^i}
\def\xjt{x_{t+1}^j}
\def\aa{\alpha}

\def\ff{\varphi}
\def\ll{\lambda}

\def\eps{\epsilon}

\def\rr{\mathbb{R}}

\newtheorem{thm}{Theorem}
\newtheorem{lm}[thm]{Proposition}

\begin{document}

\title{Stability and synchronism of certain coupled Dynamical Systems}

\author{Jos\' e A. Barrionuevo \\ Jacques A. L. Silva \\
Department of Mathematics \\ Universidade Federal RS, Brasil}

\date{}

\maketitle

\begin{abstract}{We obtain sufficient conditions for the
stability of the synchronized solution for certain classes of
coupled dynamical systems. This is accomplished by finding an
analytic expression for the transverse Lyapunov exponent through
spectral analysis. We then indicate some applications to population
dynamics. }
\end{abstract}

\section{Introduction}

The study of coupled dynamical systems has received considerable attention recently
for its interest from the mathematical, physical and biological point of view, see for
instance \cite{pikovsky}, \cite{Allen-al}, \cite{Earn-al}, \cite{TK}, \cite{Rux} and
\cite{Jost} among others. One concern about such systems is to whether or not they
will present synchronization phenomena and to whether or not such synchronization is
stable.

The system below describes the evolution of a system consisting of $d$ identical
subsystems where, on  every iteration, each subsystem undergoes its common local
evolution determined by $f$, followed by a density dependent coupling process encoded
by $C$ and $\ff$. The system can model a population consisting of $d$ patches, $x_j,\,
j=1,\dots ,d$, where, in the absence of migration, patch $j$ is controlled by a {\em
local dynamics} $\xjt = f(\xj )$. When migration is present, $\ff (f(\xj ))$
individuals leave patch $j$ and are distributed with density $c_{j i}$ on patch $i$.
The {\em global dynamics} is then
\begin{equation} \label{sist}
\xjt  = f(\xj ) - \ff (f(\xj )) + \sum_{i=1}^d c_{j i} \ff (f(\xit ));\:\: i, j =
1,\cdots , d.
\end{equation} Here $f$ is a $C^1$ map on $[0,\infty )$, $C = [c_{i j}]$ is
doubly stochastic, that is, $c_{ij} \geq 0$ and $\dd\forall i, j, \sum_{i=1}^d c_{i j}
= \sum_{j=1}^d c_{i j} = 1$. Furthermore we will assume $\ff$ differentiable a. e.
with $\ff^{\prime}$ bounded. The models in population dynamics considered in
\cite{Allen-al}, \cite{Earn-al}, \cite{GS}, \cite{SCJ1}, \cite{HD}, \cite{Rux} are all
particular cases of (\ref{sist}) for special choices of $C$ and $\ff$.

The condition on $C$ being doubly stochastic reflects that there are no losses during
the migration process. It is also necessary for the invariance of the diagonal of the
phase space, that is, for $\xj = \xit = x_t$ to be a solution of (\ref{sist}), where
each $\xj$ satisfies $\xjt = f(\xj )$.

In this paper we obtain sufficient conditions for the stability of the aforementioned
synchronized solutions. The criteria involve the Lyapunov exponents of the one
dimensional map  $f$ and of the codimension one transverse dynamical system to be
defined below.

The paper is organized as follows. In the next section we provide a criterion for
stability for general systems. In $\S$3 we improve our result for the case of normal
operators. In $\S$4, we consider a system where the coupling/migration process is time
dependent. We formulate and prove the corresponding results of the previous sections
in this setting. In the last section we indicate some applications to population
dynamics.

Previous results treat only the case where $\ff^{\prime}$ is a constant, or a 2-valued
step function as well as particular examples of matrices $C$ and in these cases we
recover such results. Our treatment only requires $\ff^{\prime}$ to be bounded and $C$
to be doubly stochastic and irreducible. This last condition is easily seen to be
necessary, see below. Moreover we are unaware of any treatment of the systems in \S 4
in the literature. Thus we extend some of the results of \cite{Earn-al}, \cite{GS},
\cite{SCJ1}, \cite{JM},\cite{Jost}, \cite{HD} and others to these more general
situations.

\section{Stability: General case}

In order to understand the behavior of orbits starting at nearby points of the
diagonal of the phase space, we first linearize (\ref{sist}). If  $\dd J_t = [\aa_{i
j}]$ denotes the Jacobian matrix of (\ref{sist}) restricted to the synchronized orbit,
we have
\[ \aa_{i j} = \left\{\begin{array}{ll}
f^{\prime}(x_t) \left( 1-(1-c_{ii})\ff^{\prime}(f(x_t))\right) , & \mbox{{\rm for }}\; i = j \\
f^{\prime}(x_t) \ff^{\prime}(f(x_t))c_{i j}, & \mbox{{\rm for }}\; i \neq j
\end{array}\right. \] We have $\dd J_t = f^{\prime}(x_t) H_t$, where $\dd H_t = I -
\ff^{\prime}(f(x_t))B$, where $B = I - C$.

We will assume that $C$ is irreducible. This is almost a necessary condition for
otherwise it permits the existence of uncoupled unsynchronized subsystems that are
each synchronized. In this case we can apply Fr\" obenius theorem \cite{House} to show
that $\ll = 1$ is the simple dominant eigenvalue of $C$, associated to the eigenvector
$v = (1,\dots ,1)$. This furnishes the decomposition $\rr^d = \rr v\oplus W$, where
$W$ is
 $C-$invariant $d-1$ dimensional subspace. Under these conditions
\begin{equation} \label{frob}
 B = P^{-1}\left[ \begin{array}{cc}
0 &  \\
  & A
\end{array}\right]P
\end{equation}
where $P$ is the matrix of the appropriate change of basis. This decomposition implies
that the stability of the synchronized solution of (\ref{sist}) is a consequence of
the stability of the trivial solution of the transversal component, $w_t$, which
satisfies
\begin{equation} \label{transv}
w_{t+1} = f^{\prime}(x_t) \left( I - \ff^{\prime}(f(x_t)) A\right) w_t.
\end{equation} We will show that under a certain integrability condition the map above is in fact
a contraction which in turn implies the stability of $w_t\equiv 0$.
The analysis of (\ref{transv}) will be based on the Lyapunov
exponents (see \cite{K-H}, \cite{Krengel}) of (\ref{transv}). Define
\[
 K_n(x) = \prod_{k=0}^{n-1} f^{\prime}(f^k(x))\left( I -
\ff^{\prime}(f^{k+1}(x))A\right) \]
 where $f^0(x)=x$ and $f^k(x)=f(f^{k-1}(x))$ for $k
> 0$. Clearly if $\mathcal{K} = \limsup ||K_n||^{1/n}$ satisfies $\mathcal{K} < 1$ we
have that (\ref{transv}) is a contraction. Now observe
\begin{equation}\label{Liap}
K_n(x) = L_n(x) \Lambda_n (x) = \left(\prod_{k=0}^{n-1}f^{\prime}(f^k(x))\right)
\left( \prod_{k=0}^{n-1}  I - \ff^{\prime}(f^{k+1}(x))A\right) \end{equation}
$L_n(x)$ depends only on the local dynamics  $f$ while $\Lambda_n(x)$ reflects also
the effects of $\ff$ and $C$. Let $\rho$ be an invariant measure of the local system.
Define for $x > 0$, $\ln^+(x) = \max (\ln (x),0)$. By Birkhoff's ergodic theorem, if
$\ln^+ |f^{\prime}|\in L^1(\rho )$, there exists $\lim_n \frac{1}{n}\sum_{k=0}^{n-1}
\ln |f^{\prime}(f^k(x))|$ for $\rho$-a.e.  $x$. For $\rho$ ergodic, this limit, call
it $L$, is independent of $x$ and is given by $\int_0^{\infty} \ln |f^{\prime}(s)|\;
d\rho (s)$. $L$ is the Lyapunov exponent of the local system governed by $f$.

Similarly the ergodic theorem of Oseledec \cite{K-H} implies that if
\[ \int_0^{\infty}
\ln^+ ||I-\ff^{\prime}(s)A|| \; d\rho (s) < \infty,\] there exists $\lim_n \frac{1}{n}
\ln ||\Lambda_n(x)|| =: \ln\Lambda (x)$ for $\rho$-a.e. $x$ and this limit is
independent of $x$ provided $\rho$ is ergodic. Our first result is
\begin{thm}\label{teo1} Consider the system  (\ref{sist}) where $f$ is a $C^1$ map,
$\ff^{\prime}$ bounded, $C$ doubly stochastic, irreducible such that $\ln^+
||I-\ff^{\prime}(s)A||$ and $\ln^+|f^{\prime}(s)|$ are in $L^1(\rho)$, where $\rho$ is
an $f-$invariant measure. Let $L = \sup_x\lim_n|L_n(x)|^{1/n}$ and $\Lambda =
\sup_x\lim_n ||\Lambda_n(x)||^{1/n}$. If $L\; \Lambda < 1$, there exists a set $E$
with $\rho (E) = 1$ such that for all $x\in E$, the synchronized solution of
(\ref{sist}) is asymptotically stable.
\end{thm}
{\em Proof:} By Oseledec's theorem there exists a set $E$ with $\rho (E) = 1$ such
that for all $x$
 in  $E$, $\lim_n ||\Lambda_n(x)||^{1/n} = \Lambda (x)$. We claim that for all $x$ in $E$
\begin{equation}\label{ineq1}
\Lambda (x) \leq \exp\left(\int_0^{\infty} \ln^+ ||I-\ff^{\prime}(s)A||\; d\rho (s)
\right)\end{equation} By the continuity of the norm and the function $\ln (\cdot )$,
the dominated convergence theorem implies that we only need to prove (\ref{ineq1}) for
$\ff^{\prime}$ simple.

Let $\ff^{\prime} (x) = \sum_{k=1}^r a_k\chi_{E_k}(x)$, where $E_k$ are measurable and
disjoint with $\rho (E_k)
> 0$ and such $E\subset\bigcup_kE_k$. For $1\leq k\leq r$, let
\[ \rho_{k,n} = \frac{\sharp\{ 0\leq j < n: f^j(x)\in E_k\}}{n}.\] This gives
\begin{eqnarray*}
\Lambda_n (x) & = & \prod_k (I - a_k A)^{n \rho_{k,n}}, \;{\mbox{\rm which imply}} \\
||\Lambda_n (x)|| & \leq & \prod_k ||I - a_k A||^{n \rho_{k,n}},\; {\mbox{\rm and}} \\
\ln ||\Lambda_n (x)|| & \leq & \sum_k {n\; \rho_{k,n}}\; \ln ||I - a_k A||
\end{eqnarray*}
Birkhoff's ergodic theorem applied
 to $\chi_{E_k}$ shows that for $1\leq k\leq r$, we have
\[ \lim_n \rho_{k,n} = \rho(E_k).\]
Therefore, given $\eps > 0$, there exists $n_0$ such that for  $n > n_0$, we have
$\rho_{k,n} \leq (1 + \eps )\rho (E_k)$. Since $\ln (\cdot )\leq \ln^+(\cdot )$,
\[
\frac{1}{n}\ln ||\Lambda_n (x)|| \leq (1+\eps )\sum_{k=1}^r \ln^+ ||I-a_kA||\; {\rho
(E_k)} = (1+\eps )\int_0^{\infty} \ln^+ ||I-\ff^{\prime}(s)A||\; d\rho (s)
\]
Note that the right hand side is independent of $x$. Thus we have
\[ ||K_n(x)||^{1/n} \leq |L_n(x)|^{1/n}\; ||\Lambda_n(x)||^{1/n} \leq (1+\eps )L\Lambda^{1+\eps}\]
and since $\eps > 0$ is arbitrary we have
\[ \lim_n ||K_n(x)||^{1/n} \leq L\;\Lambda < 1\] and thus the transversal
map (\ref{transv}) is a contraction. $\square$

\subsection{Remarks:}
\begin{description}
\item[(i)] if $E = [0,\infty )$, $L\; \Lambda \leq 1$ is necessary for the stability
of the synchronized solution.
\item[(ii)]  in certain cases, which include the case $A$ semi simple,  the norm \\
$||I-\ff^{\prime}(f^k(x))A||$ is simply the spectral radius, $\sigma_{-1}(H_t)$, of
the restriction of $B$ to the subspace $W$. Therefore, with the same hypothesis on
$f$, $\ff$ and $C$ we have
 \[ \Lambda\leq \Lambda_1 = \int_0^{\infty} \sigma_{-1}(H_{\ff^{\prime}(s)})\; d\rho (s)\]
and thus  $L\Lambda_1 < 1$ is a sufficient condition for the stability of synchronized
solution of (\ref{sist}).
\end{description}

\section{Stability: normal operators}
In the case of normal operators, that is, $AA^{\ast} = A^{\ast}A$,  one can improve
the previous result with the help of the  functional calculus of \cite{Kato} and
\cite{Rudin}. The proof of Oseledec's theorem in \cite{Ru} shows that $\lim_n
(\Lambda_n^{\ast}\Lambda_n(x))^{\frac{1}{2n}} = \Lambda (x)$ in operator norm.
Spectral analysis of $\Lambda (x)$ determines the Lyapunov exponents and respective
subspaces. Even though in our case $C$ represents a $d\times d$ matrix, the discussion
below is valid for any bounded normal operator in a Hilbert space. We present the
results in this generality since it will be applied in the future to systems more
general than (\ref{sist}). We will show that for $A$ normal, we have
\begin{equation} \label{L}
\Lambda (x) = \exp\left( \int_0^{\infty} \ln |I-\ff^{\prime}(s)A|\; d\rho (s) \right)
\end{equation}
Without loss of generality we can assume that for $s$ in $E$, $I-\ff^{\prime}(s)A$ is
nonsingular. This implies that for all $\ll\in \sigma (A)$, $\ln
|1-\ll\ff^{\prime}(s)|$ is continuous on $E$. The spectral theorem for bounded normal
operators (Theorem 12.23 in \cite{Rudin}),  allow us to define $|I-\ff^{\prime}(s)A|$
and $\ln |I - \ll\ff^{\prime}(s)A|$  by
\begin{eqnarray*}
A  & = & \int_{\sigma(A)} \ll\; dP(\ll ) \\
\ln |I-\ff^{\prime}(s)A| & = & \int_{\sigma (A)} \ln | 1 - \ll \ff^{\prime}(s)|\;
dP(\ll )
\end{eqnarray*}
where $dP(\ll )$ are the spectral projections associated with $A$. Our hypothesis on
the spectrum of $I-\ff^{\prime}(s)A$ together with Fubini's theorem imply
\[
\int_0^{\infty} \ln |I-\ff^{\prime}(s)A|\; d\rho (s)  =  \int_{\sigma (A)} \left(
\int_0^{\infty} \ln |1-\ll \ff^{\prime}(s)|\; d\rho (s)\right) \; dP(\ll )\] and since
this last integral defines a bounded operator, we take exponentials to obtain
\[ \exp\left( \int_0^{\infty} \ln
|I-\ff^{\prime}(s)A|\; d\rho (s)\right)  = \exp\left( \int_{\sigma (A)} \left(
\int_0^{\infty} \ln |1-\ll \ff^{\prime}(s)|\; d\rho (s)\right) \; dP(\ll )\right)
\] and thus the right hand side of (\ref{L}) is well defined. The spectral mapping theorem
then implies
\[
\sigma\left( \exp\left( \int_0^{\infty} \ln |I-\ff^{\prime}(s)A|\; d\rho
(s)\right)\right)  =  \left\{ \exp\left(\int_0^{\infty} \ln |1-\ll\ff^{\prime}(s) |\;
d\rho (s) \right): \ll \in \sigma (A)\right\}
\]
Our next result is
\begin{thm}\label{teo2}
Let $f$, $\ff^{\prime}$, $C$, and $L$ be as in Theorem \ref{teo1}. Let $\Lambda$ be
the spectral radius of (\ref{L}). Then, if $L\; \Lambda < 1$, the synchronized
solution of (\ref{sist}) is asymptotically stable.
\end{thm}
{\em Proof:} The above considerations imply that in order to prove the theorem we only
need to establish (\ref{L}). We will do so first assuming $\ff^{\prime}$ simple. Then
a standard limit process extends the result to $\ff^{\prime}$ in $L^{\infty}$. Let
$\ff^{\prime} = \sum_{k=1}^N \ff_k\chi_{E_k}$ where $E_k$ are disjoint measurable.
Note that since $\ff^{\prime}$ is bounded, the integral in (\ref{L}) is well defined.
As before we define $\dd\rho_{k,n} = \frac{\sharp\{ 0\leq j < n: f^j(x)\in E_k\}}{n}$.
Then
\[ (\Lambda_n^{\ast}\Lambda_n)^{\frac{1}{2n}} = \prod_{k=1}^N |I - \ff_kA|^{\rho_{k,n}}
 = \exp\left( \sum_{k=1}^N \rho_{k,n}\ln |I - \ff_kA|\right) \] Since for all $k$,
$\lim_n \rho_{n,k} = \rho (E_k)$, the continuity of $\exp (.)$ and the Lebesgue
dominated convergence theorem imply
\[ \lim_n\; \exp\left( \sum_{k=1}^N \rho_{k,n}\ln |I - \ff_kA|\right) =
\exp\left(\; \int_0^{\infty} \ln |I-\ff^{\prime}(s)A|\; d\rho (s) \right) .\] proving
(\ref{L}). This finishes the proof of Theorem \ref{teo2}.$\square$

\subsection{Remarks:}
\begin{description}
\item[(i)] Since for all $\ll\in\sigma (A)$, $|1 - \ll\ff^{\prime}(s)| \leq ||I -
\ff^{\prime}(s)A||$, we have  $\dd\Lambda  \leq \exp\left(
\int_0^{\infty}\ln^+||I - \ff^{\prime}(s)A||\; d\rho (s)\right)$,
therefore Theorem \ref{teo2} is an improvement of Theorem
\ref{teo1}.
\item[(ii)] Note that all the effects of the weighted
network relevant to synchronization are reflected on the spectrum of $\Lambda$ and can
therefore, in certain cases, be independent of $d$. It would be interesting to extend
such results to the $d = \infty$ case since Fr\" obenius theorem as well as Oseledec's
ergodic theorem are not valid without further assumptions.
\end{description}

\section{Extensions}
The analysis above can be applied to more general systems. One such
instance is the case below where the coupling/migration process no
longer depends on the density but instead obeys a seasonal dynamics,
that is, on each cycle a time dependent fraction $\mu_t$ of each
patch will mix according to a time dependent distribution $C_t =
[c_{i j}^t]$ as follows
\begin{equation} \label{sist2}
\xjt  = f(\xj ) - \mu_t\; f(\xj ) + \sum_{i=1}^d c^t_{j i}\; \mu_t\; f(\xit );\:\: i,
j = 1,\cdots , d.
\end{equation} Again $f$ is a $C^1$ map on $[0,\infty )$. For $g$ continuous on
 $[0,1]$ and $\mu_0$ arbitrary,
we assume that $\mu_t = \mu_t(\mu_0)$ is given by $\mu_{t+1} = g(\mu_t)$ . Similarly
let $\{ C_s\}_{s\in [0,1]}$ be a family of doubly stochastic irreducible matrices. If
$h$ is a continuous map on $[0,1]$ and $s_0$ is arbitrary, define $s_{t+1} = h(s_t)$
and $C_t =
 C_{t}(s_0)$ by $C_{t+1} = C_{h(s_t)}$. Assume that $G(\mu ,s) =  \mu\; C_s$ is a
measurable operator valued map with respect to the product measure $\nu\times\eta$ on
$[0,1]\times [0,1]$, where $\nu$ and $\eta$ are, respectively, a $g$-invariant and
$h$-invariant ergodic measures on $[0,1]$. Under these assumptions the diagonal of the
phase space, $\xit = \xjt = x_t$, is a synchronized solution and we are interested in
its stability. This system falls under the general theory of Random Dynamical Systems
as presented in \cite{Arnold} and thus obey, under the appropriate integrability
condition, a mutiplicative ergodic theorem. Our approach is direct and avoid the use
of this theory. Moreover, due to the particular nature of (\ref{sist2}), we are able,
as in the previous sections, to obtain much more precise information about the limit
operators than is given by the ergodic theorem alone.

The Jacobian matrix of (\ref{sist2}), $J_t = [\aa_{i j}^t]$ is now given by
\[ \aa_{i j} = \left\{\begin{array}{ll}
f^{\prime}(x_t) \left( 1-\mu_t(1-c_{ii}^t)\right) , & \mbox{{\rm for }}\; i = j \\
f^{\prime}(x_t)\;\mu_t\;c_{i j}^t, & \mbox{{\rm for }}\; i \neq j
\end{array}\right. \] yielding $J_t = f^{\prime}(x_t)(I - \mu_t B_t)$ where $B_t = I -
C_t$. Our first task is to decompose the linearized system into diagonal and
transversal components and then estimate the respective Lyapunov exponents by the
appropriate ergodic theorem. This cannot be done in general for arbitrary families $\{
C_t\}$, therefore we will consider special cases of increasing generality.
\subsection{Simultaneous Diagonalizable Matrices} This include the families of
commuting symmetric matrices and of circulant matrices. In this case there exists a
matrix $P$ such that for all $t$, $B_t = P^{-1}\; M_t\; P$ where $M_t$ is a diagonal
matrix with entries $\{ 1,\ll_2(t) ,\dots ,\ll_d(t)\}$ in its diagonal. Each
$\ll_j(t)$ is measurable and satisfies $|\ll_j(t)|\leq 1$ and $\ll_j(t)\neq 1$ by Fr\"
obenius theorem. In the symmetric (commuting) case the $\ll_j(t)$ are real and lie in
$(-1,1)$. We obtain that the transversal component of (\ref{sist2}) satisfies
\begin{equation} \label{transv2}
w_{t+1} = f^{\prime}(x_t) \left( I - \mu_t\; D_t\right) w_t.
\end{equation} where $D_t$ is the $(d-1)$ dimensional $\{\ll_2(t),\dots , \ll_d(t)\}$
diagonal matrix. As before, the synchronized solution of (\ref{sist2}) is stable if
and only if $w_t = 0$ is a stable solution of (\ref{transv2}). For each $x,\,\mu_0,\,
s_0$ in $[0,1]$, define $ L_n(x)$ as in (\ref{Liap}),  and for $j = 2,\dots ,d$,
define
\[ \Lambda_{j\, n}(\mu_0,\,s_0) = \prod_{k=0}^{n-1}  |1 - g^k(\mu_0)\,
\ll_j(h^k(s_0))|\]
The analogue of Theorem \ref{teo2} is
\begin{thm}\label{teo3} Consider the system  (\ref{sist2}) under the above conditions.
Assume in addition
\begin{description}
\item[(i)] $\ln^+|f^{\prime}(s)|$ belongs to $L^1(\rho)$, where
$\rho$ is a $f-$invariant measure on $[0,\infty )$.
\item[(ii)] For all $2\leq j\leq d$, $\ln^+|1 - \mu\,\ll_j(s)|$ belong to $L^1(\nu\times\eta )$.
\end{description}
Then there exist  sets $E\subset [0,\infty )$ and $F\subset [0,1]\times [0,1]$ with
$\rho (E) = (\nu\times\eta )\, (F) = 1$, such that for all $x\in E$ and
$(\mu_0,s_0)\in F$, the limits $L(x) = \lim_n|L_n(x)|^{1/n}$ and $\Lambda_j
(\mu_0,s_0) = \lim_n|\Lambda_{j\, n} (\mu_0,s_0)|^{1/n}$ exist. Moreover if $L =
\sup_x L(x)$ and $\Lambda = \sup_{j,\mu_0,s_0} \Lambda_j(\mu_0,s_0)$ satisfy $L\;
\Lambda < 1$, the synchronized solution of (\ref{sist2}) is asymptotically stable.
\end{thm}
{\em Proof:} The existence of $L(x)$ for $x$ in a set of full measure follows, as
before, from Birkhoff's ergodic theorem applied to $\ln L_n(\cdot )$ and (i). In
addition the limit is independent of $x$ provided $\rho$ is ergodic. The decomposition
(\ref{transv2}) allow us to derive the transversal component directly without making
use of Oseledec's theorem. For $2\leq j\leq d$, $(\mu_0, s_0)\in [0,1]\times [0,1]$,
write
\[ \Lambda_{j\, n}(\mu_0, s_0) = \exp\left( \frac{1}{n}\, \sum_{k=0}^{n-1} \ln |1 - g^k(\mu_0)\,
\ll_j(h^k(s_0))|\right)\] Condition (ii) and a $(d-1)$-fold application of Birkhoff's
ergodic theorem imply the existence of a set $F$, with $\nu\times\eta\, (F) = 1$ such
that for all $j$, all $(\mu_0,s_0)\in F$, there exists $\Lambda_j(\mu_0,s_0) = \lim_n
\Lambda_{j\, n}(\mu_0,s_0)$. If, in addition, $\nu\times\eta$ is ergodic, the limit is
independent of $(\mu_0,s_0)$ and is given by
\begin{equation} \label{L2}
\Lambda_j = \exp\left(\int_{[0,1]\times [0,1]} \ln | 1 - \mu\,
\ll_j(s)|\; d(\nu\times\eta) (\mu ,s)\right) \end{equation}  This
gives $ diag\{ \Lambda_2,\dots ,\Lambda_d\}$ as the analogue of
(\ref{L}). Its spectral radius is then $\Lambda = \max_j\Lambda_j$.
Thus, if $L\,\Lambda < 1$, the synchronized solution is
asymptotically stable.$\square$

\subsection{Symmetric (non commuting) Matrices} In this case we no longer have a
decomposition yielding a diagonal transversal component, like
(\ref{transv2}), and therefore our conclusions are somewhat weaker
than the previous theorem. Since each $C_t$ is symmetric, doubly
stochastic and irreducible, $\ll = 1$ is the dominant eigenvalue
with corresponding eigenvector $v = \{ 1, \dots ,1\}$. Moreover the
$(d-1)-$dimensional subspace $W = (\rr v)^{\bot}$ is invariant for
all $C_t$. This provides a decomposition like (\ref{frob})
\begin{equation} \label{frob2}
 B_t = P^{-1}\left[ \begin{array}{cc}
0 &  \\
  & A_t
\end{array}\right]P
\end{equation}
Accordingly the transversal component of (\ref{sist2}) now satisfies
\begin{equation} \label{transv3}
w_{t+1} = f^{\prime}(x_t) \left( I - \mu_t\; A_t\right) w_t
\end{equation}
The following theorem is the analogue of Theorem \ref{teo1} in the present situation
\begin{thm} \label{teo4}
Consider the system (\ref{sist2}) where $f$ and $\mu_t$ are as in Theorem \ref{teo3}
and assume $\{ C_t\}$ as above. For each $(\mu_0,s_0)$, define
\[ \Lambda_n(\mu_0,s_0) = \prod_{k=0}^{n-1} \left( I - g^k(\mu_0) A_{h^k(s_0)}\right) \]
Assume that $\ln^+||I - \mu\, A_{h(s)}||$ belongs to
$L^1(\nu\times\eta )$. Then there exists a set $F\subset [0,1]\times
[0,1]$ of full $\nu\times \eta$-measure such that for all
$(\mu_0,s_0)\in F$, the limit $\Lambda (\mu_0,s_0) =
\lim_n||\Lambda_n(\mu_0,s_0)||^{1/n}$ exists. Moreover if $\Lambda =
\sup_{(\mu_0,s_0)}\Lambda (\mu_0,s_0)$ satisfies $L\, \Lambda < 1$,
the synchronized solution of (\ref{sist2}) is asymptotically stable.
\end{thm}
{\em Proof:} Consider the map $\mathcal{Q}$ on $[0,1]\times [0,1]$, given by
$\mathcal{Q}(\mu,s) = (g(\mu ), h(s))$. Clearly $\nu\times\eta$ is $\mathcal{Q}$
invariant and by assumption, $\Lambda_n(\mu ,s)$ above defines a $(\nu\times\eta
)$-measurable cocycle. The existence of $\Lambda (\mu_0,s_0)$ for $(\mu_0,s_0)$ on a
set of full measure then follows from Oseledec's ergodic theorem applied to
$\Lambda_n(\mu ,s)$. We also have that for a.e. $(\mu_0,s_0)$
\begin{equation} \label{ineq2}
 \Lambda (\mu_0,s_0) \leq \exp\left( \int_{[0,1]\times [0,1]} \ln^+|| I - \mu
A_{h(s)}||\; d(\nu\times\eta )(\mu,s)\right)
\end{equation}
The proof of (\ref{ineq2}) follows the same path as the proof of
(\ref{ineq1}), thus we will omit the details. Note that since each
$A_t$ is symmetric, there exist $\ll (t)\in\sigma (A_t)$ for which
$||I - \mu\, A_t|| = |1 - \mu\ll (t)|$. $\ll (t)$ is the lowest
eigenvalue of $C_t$. If, in addition, the measure $\nu\times\eta$ is
ergodic then $\Lambda (\mu_0,s_0)$ is independent of $(\mu_0,s_0)$
and in this case we have
\[ \Lambda = \exp\left( \int_{[0,1]\times [0,1]} \ln\, |1 - \mu
\, \ll(h(s))|\; d(\nu\times\eta )(\mu,s)\right)\] as a consequence of Birkhoff's
theorem. As in the previous theorems, the condition $L\,\Lambda < 1$ implies the
asymptotic stability of the synchronized solution. $\square$
\subsection{Remarks:}
\begin{description}
\item[(i)] The assumptions on the continuity of $g$ and $h$ above
can be relaxed. All that is needed is that each map possesses an
invariant probability measure, $\nu$ and $\eta$.
\item[(ii)] Similarly to the density dependent migration, the
results above {\em do not} extend to the $d = \infty$ case even
though once again the effect of $d$ is encoded in the joint spectrum
of $\{C_t\}$.
\end{description}
\subsection{Examples:} Some special cases of (\ref{L2})
and (\ref{ineq2}) are worth mentioning.
\begin{description}
\item[(i)] if $s_0$ is a periodic point for $h$, that is, $h^p(s_0) = s_0$,
and we take $\eta = \frac{1}{p}\sum_{k=0}^{p-1} \delta_{h^k(s_0)}$,
then (\ref{L2}) becomes
\[
\Lambda_j = \exp\left(\frac{1}{p}\sum_{k=0}^{p-1}\,\int_{[0,1]} \ln
| 1 - \mu\, \ll_j(s_k)|\; d \nu (\mu )\right) \] A similar
expression corresponding to the case of a periodic point for $g$
also holds.
\item[(ii)] if $g = h$, and $\eta = \nu$, but $\mu_0\neq s_0$, we get for (\ref{L2})
\[
\Lambda_j = \exp\left(\int_{[0,1]\times [0,1]} \ln | 1 - \mu\,
\ll_j(s)|\; d(\nu\times\nu ) (\mu ,s)\right)\]
\item[(iii)] if in (ii) we have, in addition, $\mu_0 = s_0$, that is, the term
$\mu_t\, A_t$ in (\ref{transv3}) is of the form $g^t(s_0)\, A_{g^t(s_0)}$, then
(\ref{L2}) becomes
\[
\Lambda_j = \exp\left(\int_{[0,1]} \ln | 1 - s\, \ll_j(s)|\; d \nu
(s)\right)\]
\end{description}
Similar corresponding expressions can be obtained, without difficulty, to represent
(\ref{ineq2}) in the special cases above. We leave the details as well as the
formulation of other special instances of (\ref{ineq2}) to the interested reader.
These formulas are useful in cases where $\nu$ and $\eta$ are known, say Lebesgue, for
they permit the exact calculations of the Lyapunov numbers of the transversal
dynamical system.

\section{Applications}

In \cite{BS} we apply the above results to several special cases of
interest in population dynamics. In this section we will restrict
ourselves to the density dependent system (\ref{sist}) and leave the
corresponding formulations of the results related to the time
dependent system (\ref{sist2}) to the interested reader.

A direct problem consists of determining stability once $f$, $\ff$
and $C$ are given. In such situations we compute the eigenvalues
$\ll_1,\dots ,\ll_d$ of $C$. Discarding the eigenvalue $1$, we
evaluate the integrals defining the spectrum of $\Lambda (x)$ above
by Birkhoff's ergodic theorem
\begin{equation} \label{spec}
\int_0^{\infty} \ln |1 - \ll\ff^{\prime}(s) |\; d\rho (s) = \lim_n
\frac{1}{n}\sum_{k=0}^{n-1} \ln |1 - \ll\ff^{\prime}(f^k(x)) |
\end{equation}  then the maximum of these values gives the spectral radius $\Lambda$
and we can then determine stability once the Lyapunov exponent for $f$ is known. We
note that in order to use (\ref{spec}) we need the $f-$invariant measure, $\rho$, to
be absolutely continuous with respect to Lebesgue measure. Such measures, called
physical or SRB (Sinai-Ruelle-Bowen) measures, are known to exist in certain important
cases, such as expansive maps, piecewise monotonic maps of the interval as well as
{\em Axiom A} diffeomorphisms, even though a recent result of Bochi-Yoccoz shows that
they are not typical in the $C^1$-topology. Further information can be found in
\cite{Viana}, \cite{Las-Y}, \cite{Li-Y} and \cite{K-H}.

One feature of Theorems 1 and 2 is their continuous dependence on $\ff^{\prime}$, that
is, $C^1$-uniform topology. This follows directly from the formulas for $\Lambda$
above. In \cite{BS} we give examples that show that this dependence is {\em not}
continuous in $\ff$ in the $C^0$-topology.

We can also make use of the above results to study an inverse problem. In this case
$f$ and $\ff$ are given and one is interested in finding a double stochastic matrix
$C$, if possible, yielding the desired stability behavior for the synchronized
solution of (\ref{sist}). In this regard, Theorem \ref{teo5} below gives a partial
result for symmetric matrices. We will make use of the following result from
\cite{kor}.
\begin{lm} \label{prop} The following are corollaries 7 and 8 from \cite{kor}.
\begin{description}
\item[(i)] If $\{ \ll_2,\dots ,\ll_d\} \in [-1/(d-1), 1]$, then there exists a symmetric
doubly stochastic matrix $C$ with $\sigma (C) = \{ 1,\ll_2,\dots ,\ll_d\}$.
\item[(ii)] If $\ll \in (-1,1]$, there exists a positive symmetric doubly stochastic matrix
$C$ such that $\ll \in \sigma (C)$.
\end{description}
\end{lm}
The proof of Proposition \ref{prop} in \cite{kor} provides algorithms to find the
matrices $C$ in (i) and (ii) above.

\begin{thm} \label{teo5}
Let $f$ and $\ff$ be as in Theorems \ref{teo1} and \ref{teo2}. Let
$L$ denote the Lyapunov exponent of
the one dimensional map $f$. For $\ll\in [0,2]$, define \\
$\dd F(\ll ) = \exp\left(\int_0^{\infty} \ln |1-\ll\ff^{\prime}(s) |\; d\rho (s)
\right),\; m = \inf F(\ll )$ and $M = \sup F(\ll )$. Then
\begin{description}
\item[(i)] if $L\, m > 1$ the synchronized solution of (\ref{sist}) is unstable for {\em all}
symmetric configurations $C$.
\item[(ii)] if $L\, M < 1$, the synchronized solution of (\ref{sist}) is stable for
{\em all} symmetric configurations $C$.
\item[(iii)] if $L \in (\frac{1}{M}, \frac{1}{m})$, it is possible to find a symmetric doubly
stochastic matrix $C$ such that the synchronized solution of (\ref{sist}) has a
prescribed stability behavior.
\end{description}
\end{thm}
{\em Proof:} Parts (i) and (ii) are direct consequences of Theorem \ref{teo2} and the
fact that $F(\ll )$ gives the spectrum of $F(A)$. To prove (iii) we first note that
any non negative symmetric doubly stochastic matrix has its spectrum contained in
$(-1,1]$. This follows from Gershgorin's theorem \cite{Num} which states that $\sigma
(C)$ is contained in the set $\{ \ll \in \mathbb{C} : \forall i, |\ll - c_{ii}| \leq
\sum_{j\neq i} |c_{ij}| \}$. The hypothesis on $C$ then imply that $\sigma (C) \subset
[-1, 1]$.  If $C$ is positive, it is not difficult to show that $-1$ cannot be an
eigenvalue of $C$. Since $B = I - C$, we have $\sigma (B)\subset [0,2)$. Theorem
\ref{teo5} now follows easily from Theorem \ref{teo2} and Proposition
\ref{prop}.$\square$

\section{Aknowledgements}

The authors would like thank the anonymous referees for carefully
reading an earlier version of this work and providing them with a
long list of improvements. We also thank A. Lopes for some helpful
conversations.

\end{document}